\documentstyle[epsfig,aps,prl]{revtex}
\def\Journal#1#2#3#4{{#1} {\bf #2}, #3 (#4)}

\def\NPB{{\rm Nucl. Phys.} B}
\def\PLB{{\rm Phys. Lett.}  B}
\def\RMP{\rm Rev. Mod. Phys.}
\def\PR{\rm Phys. Rev.}
\def\PRL{\rm Phys. Rev. Lett.}
\def\PRD{{\rm Phys. Rev.} D}
\def\PRC{{\rm Phys. Rev.} C}

\begin{document}

\draft

\title{New empirical fits to the proton electromagnetic form factors}

\author{
E.J.~Brash,
A.~Kozlov,
Sh.~Li,
G.M.~Huber
}
\address{
\vspace{0.1cm}
University of Regina, Regina, Saskatchewan, Canada, S4S 0A2\\
}

\vspace{0.1cm}

\date{\today}

\maketitle

\begin{abstract}
Recent measurements of the ratio of the elastic electromagnetic form
factors of the proton, G$_{Ep}$/G$_{Mp}$, using the polarization
transfer technique at Jefferson Lab show that this ratio decreases
dramatically with increasing Q$^2$, in contradiction to previous
measurements using the Rosenbluth separation technique.  Using this
new high quality data as a constraint, we have reanalyzed most of the
world $ep$ elastic cross section data.  In this paper, we present a
new empirical fit to the reanalyzed data for the proton elastic magnetic
form factor in the region 0 $<$ $Q^2$ $<$ 30~GeV$^2$.  As well, we present
an empirical fit to the proton electromagnetic form factor ratio, G$_{Ep}$/G$_{Mp}$,
which is valid in the region 0.1 $<$ $Q^2$ $<$ 6~GeV$^2$.
\end{abstract}

\bibliographystyle{unsrt}


The elastic electromagnetic form factors are crucial to our
understanding of the proton's internal structure.  Indeed, the
differential cross section for elastic $ep \rightarrow ep$ scattering
is described completely in terms of the Dirac and Pauli form factors,
$F_1$ and $F_2$, respectively, based solely on fundamental symmetry
arguments.  Further, the Sachs form factors, $G_{Ep}$ and $G_{Mp}$,
which are simply derived from $F_1$ and $F_2$, reflect the
distributions of charge and magnetization current within the proton.

Until recently, the form factors of the proton have been determined
experimentally using the Rosenbluth separation
method~\cite{Rosenbluth}, in which one measures elastic $ep$ cross
sections at constant $Q^2$, and varies both the beam energy and
scattering angle to separate the electric and magnetic contributions.
In terms of the Sachs form factors, the differential cross section for
elastic $ep$ scattering has traditionally been written as
\begin{equation}
\label{eq::cross}
\frac{d\sigma}{d\Omega} = \sigma_{ns} \left( \frac{G_{Ep}^2 + \tau
G_{Mp}^2}{1+\tau} + 2\tau G_{Mp}^2 \tan^2(\theta_e/2) \right),
\end{equation}
where $\tau=Q^2/4M_p^2$, $\theta_e$ is the in-plane electron scattering angle.
For elastic $ep$ scattering, the so-called nonstructure cross 
section, $\sigma_{ns}$ is given by
\begin{equation}
\label{eq:sigmans}
        \sigma_{ns} = \frac{ \alpha_{em}^2 \cos^2(\theta_e/2) } 
{4E^2 \sin^4 (\theta_e/2)}
                      \frac{E^\prime}{E}
\end{equation}
where $\alpha_{em}$ is the electromagnetic coupling constant, and
$E^\prime$($E$) is the energy of the scattered (incident) electron.

From the measured differential cross section, 
one typically derives a ``reduced cross section'', defined
according to
\begin{equation}
\label{eq::rosenbluth}
\sigma_R = \frac{d\sigma}{d\Omega} 
\frac{(1+\tau)\epsilon}{\sigma_{ns}\tau}
= \frac{\epsilon}{\tau}G_{Ep}^2(Q^2) + G_{Mp}^2(Q^2),
\end{equation}
where $\epsilon=\{1+2(1+\tau)\tan^2(\theta_e / 2)\}^{-1}$ is a measure of
the virtual photon polarization.  Equation~\ref{eq::rosenbluth} is
known as the Rosenbluth formula, and shows that fits to reduced cross
section measurements made at constant $Q^2$ but varying $\epsilon$
values may be used to extract both form factors independently.

\begin{figure}[htb]
\begin{center}
\epsfig{file=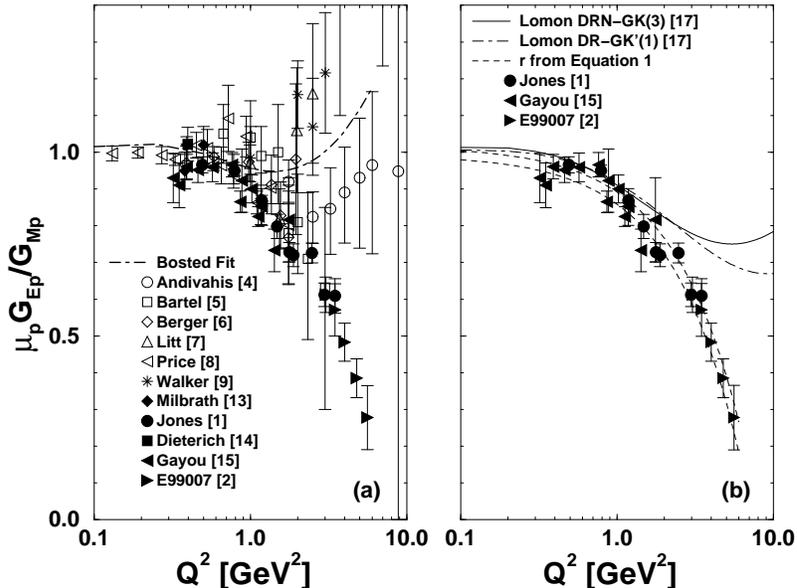,height=4.0in,angle=-90}
\end{center}
\caption[]{(a) Published world data for $r = \mu_p G_{Ep}/G_{Mp}$; 
open symbols indicate Rosenbluth 
separations~\cite{Andivahis,Bartel,Berger,Litt,Price,Walker}
while filled symbols indicate polarization transfer 
measurements~\cite{Jones,E99007,Milbrath,Dieterich,Gayou}. In the
case of the Rosenbluth data, the error bars shown are the result of
combining systematic and statistical errors in quadrature.  For the
polarization transfer measurements, the error bars shown are statistical
only; the systematic uncertainties are in general a small fraction of 
the statistical error in these experiments.
The dot dash line is the parameterization from ref.~\cite{Bostedparam} 
to the cross section data, which indicates $r$ $\approx$ $1$.
(b) Fit to polarization transfer measurements from Jefferson Lab.  Included
are the most recent data at large $Q^2$ from E99007~\cite{E99007}.  Also
shown are calculated ratios from recent fits to the electromagnetic form
factors by Lomon~\cite{Lomon} within the Gari-Kr\"umpelmann framework.}
\label{fig:gepgmp}
\end{figure}

With increasing $Q^2$, the reduced cross sections are increasingly
dominated by the magnetic term $G_{Mp}$; at $Q^2$ $\approx$ 3~GeV$^2$,
the electric term contributes only a few percent of the cross section.
Furthermore, referring to the open data points in the left panel of
Fig.~\ref{fig:gepgmp}, one can see that the various Rosenbluth
separation data
sets~\cite{Andivahis,Bartel,Berger,Litt,Price,Walker} for the
ratio $\mu_p G_{Ep}/G_{Mp}$, where $\mu_p=2.79$ is the magnetic
moment of the proton, are not consistent with one another for
$Q^2$ $>$ 1~GeV$^2$.  It is clear that a tremendous effort has gone
into the analysis of these difficult experiments, however, one is
forced to speculate that some of the experiments have underestimated
the systematic errors.  For example, the Rosenbluth experiments apply
radiative corrections to their data at leading order, i.e. one hard
photon emitted, which can vary significantly with $\epsilon$ depending
on the specific experiment.  However, higher order radiative
corrections, i.e. involving more than one photon, could in fact change
the slope of the reduced cross section versus $\epsilon$ plot, and
thus have an impact on the extracted form factor ratio.

Due to the fundamental nature of the quantities at hand, a more robust
method for measuring the proton electromagnetic form factors is
certainly desirable. Over the last few years, focal plane polarimeters
have been installed in hadron spectrometers in experimental facilities
at Bates, Mainz, and Jefferson Lab.  Specifically, one makes use of
the polarization transfer method~\cite{Rekalo,Arnold}, in which one
measures, using a focal plane polarimeter, the transverse ($P_t$) and
longitudinal ($P_\ell$) components of the recoil proton polarization
in $^1H(\vec{e},e^\prime \vec{p})$ scattering, using a longitudinally
polarized electron beam.

The proton form factor ratio is given simply by
\begin{equation}
r = \mu_p {G_{Ep} \over G_{Mp}} = 
- {P_t \over P_\ell} { \mu_p \left (E_e + E_{e^\prime } \right ) \over 2M_p } 
\tan \left (\theta _e/2 \right ).
\label{eq:ratio}
\end{equation}
Here, $E_e$ ($E_{e^\prime}$) is the incident (scattered) electron energy.
The polarization transfer method offers a number of advantages over
the traditional Rosenbluth separation technique.  Using the ratio of
the two simultaneously measured polarization components greatly
reduces systematic uncertainties.  For example, a detailed knowledge
of the spectrometer acceptances, something which plagues the cross
section measurements, is in general not needed.  Moreover, it is not
necessary to know either the beam polarization or the polarimeter
analyzing power, since both of these quantities cancel in measuring
the ratio of the form factors.  The dominant systematic uncertainty is
the knowledge of spin transport, although in comparison to the size of
the systematic uncertainties in cross section measurements, this too
is small.  Finally, the most recent theoretical calculations~\cite{Afanasev}
 of the effects of radiative processes verify the assertion that the
form factor ratio is unaffected at the momentum transfers involved here.  

As mentioned above, the proton form factor ratio has been measured at
several facilities using the polarization transfer technique.  It was
first used by Milbrath {\it et al}.~\cite{Milbrath}, who determined
$r$ at $Q^2$ = 0.38 and 0.50~GeV$^2$, and the result was in good
agreement with both the Rosenbluth separation results and a subsequent
polarization measurement at $Q^2$ = 0.4 by Dieterich {\it et
al}.~\cite{Dieterich} at Mainz.  However, polarization transfer
measurements up to $Q^2$=3.5~GeV$^2$ in Hall A at Jefferson
Lab~\cite{Jones,Gayou} have revealed the somewhat surprising result
shown in Fig.~\ref{fig:gepgmp}, that the form factor ratio decreases
with increasing $Q^2$.  Most recently, this trend has been confirmed
in Jefferson Lab Experiment E99007~\cite{E99007}, which extends the form factor
ratio measurement to $Q^2$=5.6~GeV$^2$; these new data are also shown
in Fig.~\ref{fig:gepgmp}.  We have fit the Jefferson Lab data using a simple 
linear parameterization, i.e.,
\begin{equation}
r = 1.0 - (0.130 \pm 0.005) \left[ Q^2 - (0.04 \pm 0.09) \right],
\label{eq:fit}
\end{equation}
for 0.04 $<$ $Q^2$ $<$ 5.6~GeV$^2$.  This empirical description, which
gives an acceptable fit to the Jefferson Lab data with the smallest
number of free parameters, is shown in the right panel of
Fig.~\ref{fig:gepgmp} using two dashed curves to represent the range
of uncertainty in the fit parameters.  Note that this description of
the ratio is singularly inconsistent with the global fit from
Ref.~\cite{Bostedparam} shown in the left panel of
Fig.~\ref{fig:gepgmp}.  However, a similar decreasing ratio has been
reported in a number of recent theoretical models.  In addition, we
also show in the right panel of Fig.~\ref{fig:gepgmp} ratios
calculated from recent fits to the electromagnetic form factors by
Lomon~\cite{Lomon} within the framework of the Gari-Kr\"umpelmann
model.  The two curves shown differ somewhat from one another in their
specific choice of the form of the hadronic form factors as well as in
the parametrization of the behaviour at large $Q^2$, and as such represent
upper and lower bounds on the extracted form factor ratio.
However, these curves both lie significantly
above the new data.

In the remainder of this paper, for the purposes of reanalyzing the
cross section data, we have used an empirical prescription for the
form factor ratio.  For $Q^2$ $<$ 0.04~GeV$^2$, we have used $r=1$;
for 0.04 $<$ $Q^2$ $<$ 7.7~GeV$^2$, we have employed Eq.~\ref{eq:fit};
for $Q^2$ $>$ 7.7~GeV$^2$, we have used $r=0$.  The boundary of 
0.04~GeV$^2$ (7.7~GeV$^2$) corresponds to the value of $Q^2$ where
Eq.~\ref{eq:fit} predicts a ratio of 1 (0).  The choice of setting
$r=0$ for the largest $Q^2$ region is somewhat arbitrary, since no
experimental data exists in this kinematic regime.  However, since the
electric contribution to the total cross section is minimal in this
$Q^2$ region, our choice has in fact little impact on the extracted
value of the magnetic form factor.  In addition, our linear fit almost
certainly has the wrong asymptotic behavior at very large $Q^2$, based
on theoretical expectations, and therefore we have extended the
empirical fit only to $Q^2$=7.7~GeV$^2$ to match the higher $Q^2$
assumption of $r=0$.

As stated above, the new Jefferson Lab data for the form factor ratio
is in general disagreement with the higher $Q^2$ Rosenbluth separation
measurements.  Since the Rosenbluth separation measurements
systematically attribute more strength in the cross section to the
electric part (larger ratio, $r$), this means that their extracted
values for the magnetic form factor, $G_{Mp}$, are potentially
systematically too small.

We have reanalyzed the available cross section data
using the following procedure:
\begin{itemize}
\item In the cases where experiments have extracted reduced cross
  section data at multiple $\epsilon$ values at each $Q^2$
  (Refs.~\cite{Andivahis,Bartel,Berger,Litt,Walker,Janssens}), we
  have reanalyzed these data using the aforementioned form factor ratio
  prescription.  The net effect of this procedure is that the form
  factor ratio constraint fixes the ratio of the slope to the intercept
  of the graph of reduced cross section vs. $\epsilon$.  Therefore, in
  practice, one extracts only a single parameter from the new fit to
  the data.
\item For the data of Sill {\it et al.}~\cite{Sill}, which presents
  reduced cross section data at a single $\epsilon$ value for each
  $Q^2$, we extract $G_{Mp}$ using the above form factor ratio
  prescription directly.  The authors had assumed, quite
  reasonably at the time, $r=1$ in their extraction of $G_{Mp}$.
  \item It is important to recognize that Eqs.~\ref{eq::cross} and
  \ref{eq:sigmans} describe the elementary single photon exchange
  electron scattering cross section, i.e., the Born cross section.
  However, other processes, such as vacuum polarization and radiative
  effects, contribute to the total cross section that one measures in
  an experiment.  The technique that has been used to date is to
  correct the measured $ep$ cross section to account for the
  contribution from these extra processes, which are fully calculable
  within the framework of quantum electrodynamics, and thus extract
  the Born cross section~\cite{MoTsai}. This being stated, the
  calculation of the vacuum polarization processes has varied between
  the different experiments.  Specifically, only the most recent
  analyses~\cite{Andivahis,Walker,Sill} have included the $\mu^+
  \mu^-$ and $q\bar{q}$ vacuum polarization contributions.  These
  processes were not accounted for in the original work of Mo and
  Tsai~\cite{MoTsai}.  As a result, in our reanalysis of the cross
  section data we have included, using the same formalism as presented
  in~\cite{Walker}, a correction to the extracted Born cross sections
  in the older experiments~\cite{Bartel,Berger,Litt,Janssens}.  The
  size of this correction is almost negligible for $Q^2$ $<$
  0.3~GeV$^2$, but increases to about 1.3\% at Q$^2$=4~GeV$^2$, which
  is the largest momentum transfer probed in the earlier measurments.
  We also point out that in the hadronic and particle physics
  communities, the vacuum polarization processes are often modelled in
  terms of a strengthening of the electromagnetic coupling, or
  alternatively in terms of a ``screening'' of the bare electron
  charge at low $Q^2$.
\end{itemize}

\begin{figure}[htb]
\begin{center}
\epsfig{file=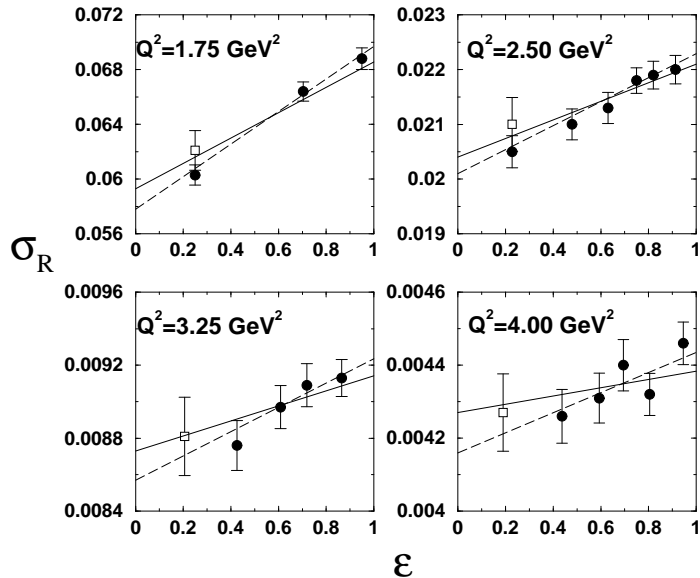,width=3.0in,angle=-90}
\end{center}
\caption[]{The reduced cross section data of Andivahis {\it et al.} for 
the four lowest $Q^2$ values of 1.75, 2.50, 3.25, and 4.00~GeV$^2$.  
The open and closed symbols in the figure are from two
different spectrometers.  The solid (dashed) line represents the
fit using the form factor ratio constraint (direct extraction method).
The details of the fitting procedure, as well as the
calculation of uncertainties
and normalization factors, are discussed in text.}
\label{fig:andi}
\end{figure}

As an example of the effect of form factor ratio constraint, in
Fig.~\ref{fig:andi}, we show reduced cross section data which has been
recalculated using the original data from Andivahis {\it et al.}  for
the four lowest $Q^2$ values of 1.75, 2.50, 3.25, and 4.00~GeV$^2$.
In each case, the dashed line is the best fit line using the direct
Rosenbluth method.  The solid line is the best fit using our form
factor ratio constraint.  The error bars shown are statistical and
point-to-point systematic errors, as reported in
Ref.~\cite{Andivahis}, added in quadrature.  The open and closed
symbols in the figure are from two different spectrometers, known as
the ``1.6~GeV'' and ``8~GeV'' spectrometers, respectively. In
Ref.~\cite{Andivahis}, the authors report that the data from these two
spectrometers were cross-normalized at the two lowest $Q^2$ points,
which resulted in a renormalization factor of $0.956$ being applied to
the 1.6~GeV spectrometer data. In this paper, the authors also report
an overall normalization uncertainty of $\delta_{norm}=1.77\%$.  We
have not included the renormalization uncertainty in each data point,
but it has been included in the final uncertainty in the intercept of
each graph, as
\begin{equation}
(\delta b_{final})^2 = (\delta b_{raw})^2 + (b_{raw} \cdot \delta_{norm})^2,
\end{equation}
where $b_{raw}$ is the intercept of the straight-line fit to the data.

We note that our new fits
would correspond to increasing the normalization factor of the 1.6~GeV
spectrometer, or alternatively introducing a momentum dependence to
the 1.6~GeV spectrometer acceptance.  In an attempt to address this issue,
we have used instead a renormalization factor of $0.980$
for the 1.6~GeV data, 
which represents an upper limit derived from the
original normalization factor and uncertainty from Ref.~\cite{Andivahis}.
 
In addition, in the final extraction of $G_{Mp}$ using the form factor
ratio constraint, we have incorporated (in quadrature) the uncertainty
in the form factor ratio itself, as expressed in Eq.~\ref{eq:fit}.
The results of the two extraction methods, including the final
uncertainties, are summarized in Table~\ref{tab:extraction}.
\begin{table}[htb] 
 \begin{center}
  \caption{Compilation of all $\frac{G_{M_p}}{\mu_p G_D}$ extraction results. 
The details of the procedure for determing uncertainties and renormalization
factors are discussed in the text.}
  \label{tab:extraction}

  \begin{tabular}{cccccc}

  $Q^2 \atop (\rm{GeV}^2)$ & Direct Extraction & New Extraction   & $Q^2 \atop (\rm{GeV}^2)$ & Direct Extraction & New Extraction   \\ \hline
   &   \multicolumn{2}{c}{Andivahis {\it et al.}\cite{Andivahis}} & &\multicolumn{2}{c}{Berger {\it et al.}\cite{Berger}  }  \\
1.750   & 1.053  $\pm$ 0.013   &   1.067  $\pm$  0.010     & 0.389   & 0.985   $\pm$ 0.027  &  0.986   $\pm$ 0.021   \\  
2.500   & 1.058  $\pm$ 0.012   &   1.066  $\pm$  0.010     & 0.584   & 0.985   $\pm$ 0.021  &  1.002   $\pm$ 0.021   \\  
3.250   & 1.053  $\pm$ 0.015   &   1.063  $\pm$  0.010     & 0.779   & 1.005   $\pm$ 0.024  &  1.015   $\pm$ 0.021   \\  
4.000   & 1.040  $\pm$ 0.015   &   1.053  $\pm$  0.010     & 0.973   & 1.005   $\pm$ 0.026  &  1.035   $\pm$ 0.022   \\  
5.000   & 1.028  $\pm$ 0.015   &   1.035  $\pm$  0.010     & 1.168   & 1.024   $\pm$ 0.031  &  1.051   $\pm$ 0.023   \\  
6.000   & 1.002  $\pm$ 0.019   &   1.012  $\pm$  0.011     & 1.363   & 1.038   $\pm$ 0.030  &  1.044   $\pm$ 0.023   \\  
7.000   & 0.973  $\pm$ 0.022   &   1.002  $\pm$  0.013     & 1.558   & 1.032   $\pm$ 0.034  &  1.062   $\pm$ 0.024   \\  
   & \multicolumn{2}{c}{Bartel {\it et al.}\cite{Bartel} } & 1.752   & 1.064   $\pm$ 0.042  &  1.066   $\pm$ 0.025   \\  
0.670   & 0.964  $\pm$ 0.027   &   1.006  $\pm$  0.014     & &  \multicolumn{2}{c}{Janssens {\it et al.}\cite{Janssens} }  \\
1.000   & 1.014  $\pm$ 0.028   &   1.056  $\pm$  0.015     & 0.156   & 0.926   $\pm$ 0.027  &  0.979   $\pm$ 0.013   \\
1.169   & 1.017  $\pm$ 0.024   &   1.052  $\pm$  0.014     & 0.179   & 0.960   $\pm$ 0.016  &  0.968   $\pm$ 0.009   \\
1.500   & 1.031  $\pm$ 0.030   &   1.065  $\pm$  0.015     & 0.195   & 1.001   $\pm$ 0.026  &  0.998   $\pm$ 0.013   \\
1.750   & 1.052  $\pm$ 0.023   &   1.052  $\pm$  0.015     & 0.234   & 0.937   $\pm$ 0.025  &  0.985   $\pm$ 0.011   \\
3.000   & 1.052  $\pm$ 0.023   &   1.052  $\pm$  0.014     & 0.273   & 0.940   $\pm$ 0.017  &  0.961   $\pm$ 0.009   \\
    &   \multicolumn{2}{c}{Litt {\it et al.}\cite{Litt} }  & 0.292   & 0.934   $\pm$ 0.020  &  0.965   $\pm$ 0.009   \\
1.500   & 0.972  $\pm$ 0.114   &   1.071  $\pm$  0.022     & 0.312   & 0.966   $\pm$ 0.016  &  0.965   $\pm$ 0.009   \\
2.000   & 0.979  $\pm$ 0.074   &   1.070  $\pm$  0.022     & 0.350   & 0.973   $\pm$ 0.025  &  0.973   $\pm$ 0.012   \\
2.500   & 1.011  $\pm$ 0.034   &   1.064  $\pm$  0.022     & 0.389   & 0.958   $\pm$ 0.014  &  0.983   $\pm$ 0.008   \\
3.750   & 0.971  $\pm$ 0.041   &   1.072  $\pm$  0.022     & 0.428   & 0.969   $\pm$ 0.024  &  0.998   $\pm$ 0.012   \\
    &   \multicolumn{2}{c}{Sill {\it et al.}\cite{Sill} }  & 0.467   & 0.976   $\pm$ 0.016  &  0.994   $\pm$ 0.009   \\
2.862   & 1.023  $\pm$ 0.018   &   1.063  $\pm$  0.021     & 0.506   & 0.957   $\pm$ 0.024  &  0.988   $\pm$ 0.012   \\
3.621   & 1.024  $\pm$ 0.020   &   1.060  $\pm$  0.023     & 0.545   & 0.986   $\pm$ 0.016  &  1.001   $\pm$ 0.009   \\
5.027   & 1.007  $\pm$ 0.018   &   1.040  $\pm$  0.019     & 0.584   & 0.982   $\pm$ 0.024  &  1.003   $\pm$ 0.012   \\
4.991   & 1.011  $\pm$ 0.019   &   1.042  $\pm$  0.021     & 0.623   & 0.989   $\pm$ 0.016  &  0.997   $\pm$ 0.008   \\
5.017   & 1.000  $\pm$ 0.018   &   1.027  $\pm$  0.019     & 0.662   & 1.028   $\pm$ 0.025  &  1.017   $\pm$ 0.012   \\
7.300   & 0.949  $\pm$ 0.019   &   0.973  $\pm$  0.020     & 0.701   & 0.984   $\pm$ 0.017  &  1.009   $\pm$ 0.009   \\
9.629   & 0.891  $\pm$ 0.019   &   0.907  $\pm$  0.020     & 0.740   & 1.017   $\pm$ 0.025  &  1.039   $\pm$ 0.012   \\
11.99   & 0.873  $\pm$ 0.019   &   0.885  $\pm$  0.020     & 0.779   & 1.037   $\pm$ 0.018  &  1.023   $\pm$ 0.009   \\
15.72   & 0.821  $\pm$ 0.026   &   0.829  $\pm$  0.025     & 0.857   & 1.086   $\pm$ 0.018  &  1.068   $\pm$ 0.011   \\
19.47   & 0.732  $\pm$ 0.028   &   0.738  $\pm$  0.029     & & \multicolumn{2}{c}{Walker {\it et al.}\cite{Walker} }  \\
23.24   & 0.729  $\pm$ 0.033   &   0.734  $\pm$  0.033     & 1.000   & 1.002   $\pm$ 0.028  &  1.045   $\pm$ 0.011  \\
26.99   & 0.710  $\pm$ 0.041   &   0.713  $\pm$  0.042     & 2.003   & 1.016   $\pm$ 0.013  &  1.076   $\pm$ 0.011  \\
31.20   & 0.721  $\pm$ 0.064   &   0.723  $\pm$  0.064     & 2.497   & 1.011   $\pm$ 0.013  &  1.075   $\pm$ 0.011  \\
        &                      &                           & 3.007   & 1.003   $\pm$ 0.013  &  1.072   $\pm$ 0.010 
  \end{tabular}
  \end{center}
\end{table}

\begin{figure}[htb]
\begin{center}
\epsfig{file=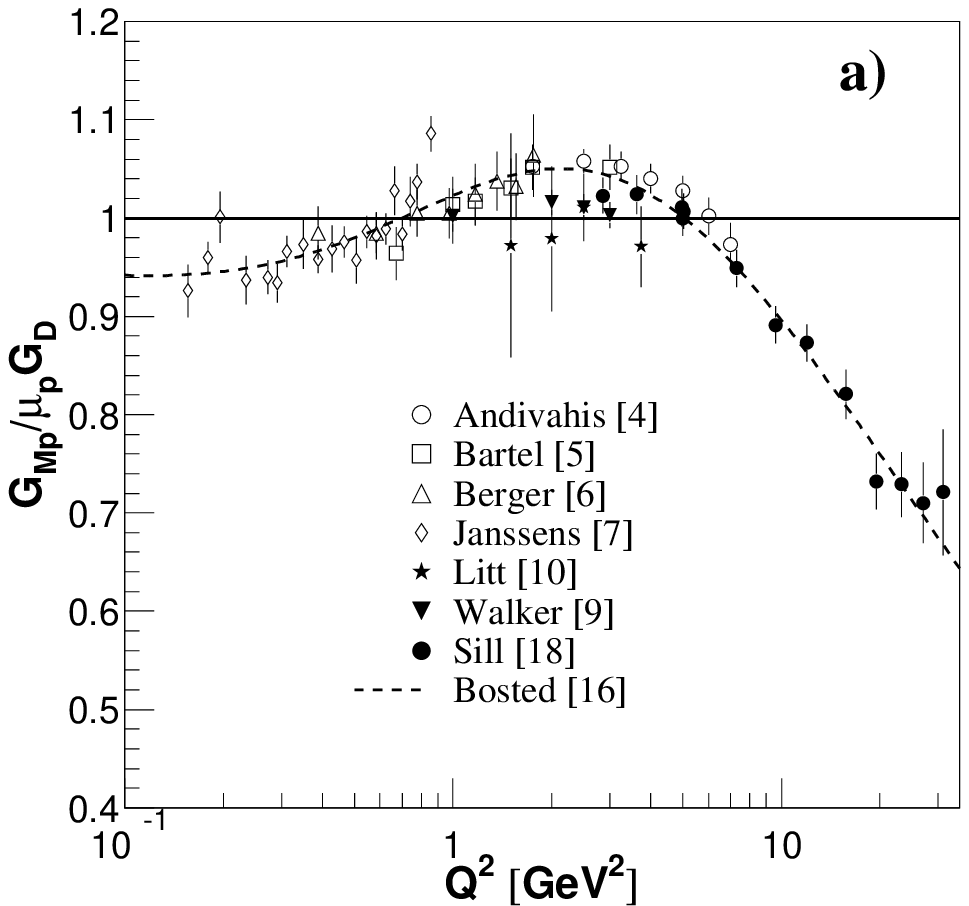,width=3.2in,angle=0}
\epsfig{file=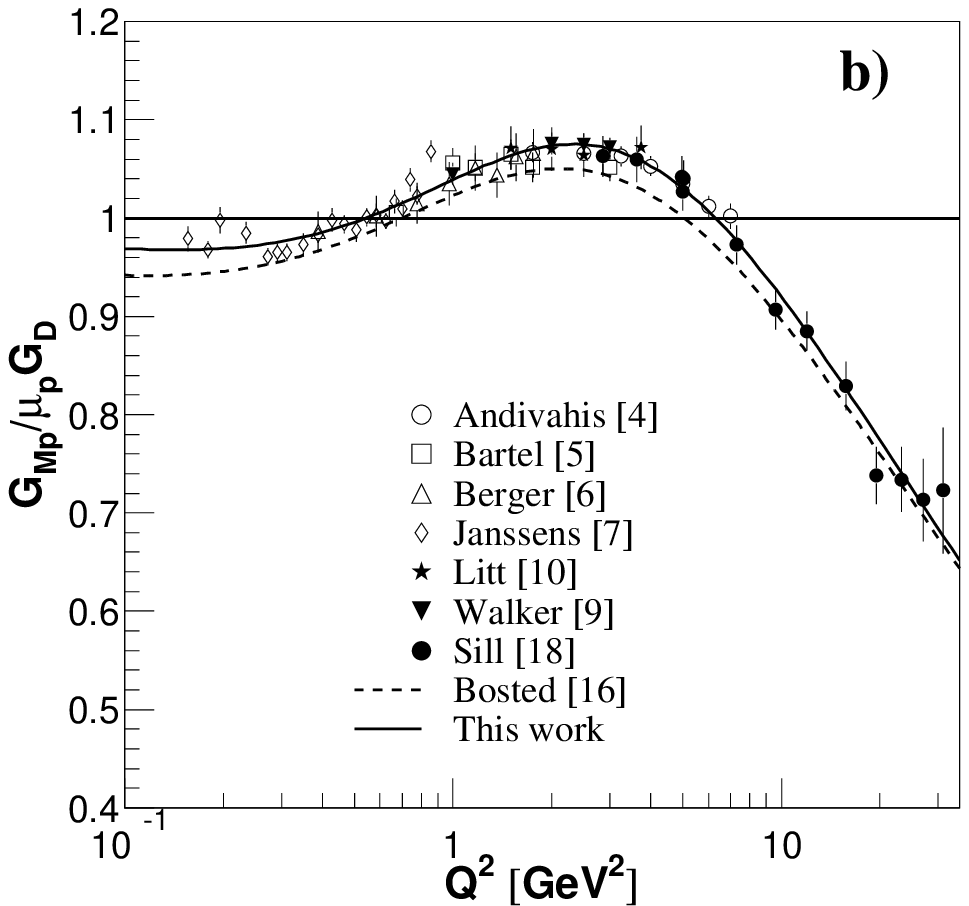,width=3.2in,angle=0}
\end{center}
\caption[]{a) The proton magnetic form factor,
expressed as $G_{Mp}/\mu_p G_D$, as published in the literature.  
b) The proton magnetic form factor, after reanalysis using the form
factor ratio constraint. The solid line is the new empirical fit,
as explained in the text, while the dashed line is the parameterization from
Ref.~\cite{Bostedparam}. 
}
\label{fig:newgmp}
\end{figure}

In Fig.~\ref{fig:newgmp}, we show data for the proton magnetic form
factor, expressed as $G_{Mp}/\mu_p G_D$, where $G_D=(1+Q^2/0.71)^{-2}$
is the dipole form factor parameterization. In the left panel, we show
the magnetic form factor as extracted using 
direct Rosenbluth separation (or using the assumption of $r=1$ in the case of
the data of Sill {\it et al.}).  In the right panel, we show the newly
extracted data using the above constraint on the form factor ratio.
In both panels, the dashed curve is the parameterization of
Bosted~\cite{Bostedparam}, while in the right panel, the solid line is our
new empirical fit, and is given by
\begin{equation}
\frac{G_{Mp}}{\mu_p} = \frac{1}{1+(0.116\pm0.040) Q + (2.874\pm0.098) Q^2 + 
(0.241\pm0.107) Q^3 + (1.006\pm0.069) Q^4
 +(0.345\pm0.017) Q^5},
\end{equation}
the form of which is consistent with Ref.\cite{Bostedparam}. 

Imposing the form factor ratio constraint has the expected effect that the
extracted magnetic form factor is systematically larger than when one
uses the direct Rosenbluth method.  Indeed, the data in the right
panel of Fig.~\ref{fig:newgmp} lie approximately 1.5-3\% above the
Bosted parameterization.  
As mentioned
earlier, the decreased electric strength implicit in a decreasing form
factor ratio results in increased magnetic strength.  However, perhaps
the most striking feature of the reanalyzed data is that imposing the
constraint on the form factor ratio results in uncertainties that are
much reduced compared to using the direct Rosenbluth separation.  This
is due simply to the fact that we are extracting only a single
parameter (proportional to $G_{Mp}$) from the cross section data, as
opposed to extracting two parameters, as is done with the Rosenbluth
method.  Also, the Bosted parameterization and our new fit
converge at large $Q^2$.  As stated previously, the electric strength
at large $Q^2$ decreases rapidly, and so indeed our choice of $r=0$,
compared to the previous choice of $r=1$, has little effect on the 
magnetic form factor extracted from the data of Sill {\it et al.} 

One also sees immediately that in using the new form factor
constraint, the extracted magnetic form factor data from the various
experiments are more consistent with one another, as well as with the
new parameterization.  Comparing the data extracted using
direct Rosenbluth separation to the Bosted parameterization, we
calculate $\chi^2/N_{d.f.}$=0.97. This is somewhat larger
than the value quoted in Ref.~\cite{Bostedparam}, since we have used a
different data sample.  Using the form factor constraint, and
comparing to our new parameterization, $\chi^2/N_{d.f.}$=0.82.  
Based on our new parameterizations of both $G_{Mp}$ and the form
factor ratio, we have calculated the elastic hydrogen differential cross
section directly using Eq.~\ref{eq::cross}.  
We find that the deviation of the cross section
calculated using the new fits from the measured world cross section
data~\cite{Andivahis,Bartel,Berger,Litt,Walker,Janssens} form an 
approximately Gaussian
distribution around -1\%, with a standard deviation of $\sim$2.5\%.
As well, there is no significant $Q^2$ dependence of 
the deviation 
in the region  0.1 $<$ $Q^2$ $<$ 30~GeV$^2$.


In conclusion, we have used recent measurements of the ratio of the 
elastic electromagnetic form factors of the proton, G$_{Ep}$/G$_{Mp}$, 
using the polarization transfer technique as a constraint in reanalyzing
most of the world $ep$ elastic cross section data.  We have presented a
new empirical fit to the reanalyzed data for the proton elastic magnetic
form factor in the region 0 $<$ $Q^2$ $<$ 30~GeV$^2$, and find that over most of
this kinematic region, the magnetic form factor is systematically 1.5-3\% 
larger than had been extracted in previous analyses. 

We thank R.L.~Lewis for useful discussions regarding the vacuum polarization
and radiative correction effects.
This work was supported by the Natural Sciences and Engineering
Research Council of Canada.

\end{document}